\documentclass[mathleft
]{an}
\usepackage{graphicx}
\usepackage{times}
\usepackage{natbib}
\begin{document}

\Pagespan{789}{}
\Yearpublication{2006}%
\Yearsubmission{2005}%
\Month{11}%
\Volume{999}%
\Issue{88}%

\title{Planet formation in post-common-envelope binaries}

\author{Dominik R.G. Schleicher\inst{1}\fnmsep\thanks{Corresponding author:
  \email{dschleic@astro.physik.uni-goettingen.de}\newline}
\and  Stefan Dreizler \inst{1}
\and Marcel V\"olschow \inst{2}
\and Robi Banerjee \inst{2}
\and Frederic V. Hessman\inst{1}
}
\titlerunning{Planet formation in post-common-envelope binaries}
\authorrunning{D.R.G. Schleicher}
\institute{
Institute f\"ur Astrophysik G\"ottingen, Friedrich-Hund-Platz 1, 37077 G\"ottingen, Germany
\and 
Hamburg Observatory,  Gojenbergsweg 112, 21029 Hamburg, Germany
}

\received{30 May 2005}
\accepted{11 Nov 2005}
\publonline{later}

\abstract{
To understand the evolution of planetary systems, it is  important to investigate planets in highly evolved stellar systems, and to explore the implications of their observed properties with respect to potential formation scenarios. Observations suggest the presence of giant planets in post-common-envelope binaries (PCEBs). A particularly well-studied system with planetary masses of $1.7$~M$_J$ and $7.0$~M$_J$ is NN Ser. We show here that a pure first-generation scenario where the planets form before the common envelope (CE) phase and the orbits evolve due to the changes in the gravitational potential is inconsistent with the current data. We  propose a second-generation scenario where the planets are formed from the material that is ejected during the CE, which may naturally explain the observed planetary masses. In addition, hybrid scenarios where the planets form before the CE and evolve due to the accretion of the ejected gas appear as a realistic possibility.}


\maketitle

\section{Introduction}
While observational studies have predominantly focused on planetary systems around single stars, it is well-known that most of the stars are in binaries. It is therefore important to understand whether planets can form in binary systems, which is suggested by an increasing amount of recent observations. The unevolved dG/dM binary Kepler 47 harbors two planets with orbital periods of $49.5$~days and $303.2$~days \citep{Orosz12}. The binary system Kepler 16 hosts the Saturn-sized planet Kepler 16b \citep{Doyle11} and Kepler 34/35 hosts a planet with $1/5$ of Jupiter's mass \citep{Welsh12}. From a theoretical point of view, the formation of planets in binary systems has been explored e.g. by \citet{Hagi07}. 

To understand the evolution of planetary systems, it is particularly relevant to search for planets in highly evolved stellar systems, to determine if and under which conditions they survive different phases of stellar evolution. The first exoplanets have  been found around pulsars, which provide accurate clocks for a precise measurement \citep[e.g.][]{Wolszczan92, Konacki03, Yan13}. Recent observations have revealed the presence of a protoplanetary debris  disk around the pulsar 4U~0142+61 \citep{Wang06, Ertan07}. Planet formation scenarios in highly evolved stellar systems have been put forward by \citet{Perets10} and \citet{Tutukov12}, considering for instance the formation of planets in gaseous disks formed by stellar winds in binary systems.

An important class of highly evolved binaries where the presence of planets is considered are the post-common envelope binaries (PCEBs). These are compact binary systems of only a few solar radii consisting of a white dwarf and a low-mass main sequence star. For the PCEB system NN Ser, different series of mid-eclipse times have been obtained since the discovery of the eclipses in 1998 \citep{Haefner04, Brinkworth06, Beuermann10, Beuermann13}. In 2010, the quality of the data was good enough to obtain a two-planet solution \citep{Beuermann10}, which was shown to be dynamically stable \citep{Beuermann13}. The most recent data suggest planetary masses of $1.7$~M$_J$ and $7.0$~M$_J$ with low eccentricities and semi-major axis of $\sim3.3$~AU and $\sim5.4$~AU \citep{Beuermann13}. The planetary solution has been  confirmed by \citet{Marsh14}. 

At the same time, there is increasing evidence against  alternative explanations of the eclipsing time variations. A frequently considered mechanism are  periodic changes in the stellar structure related to  magnetic activity, which could give rise to regular changes in the star's quadrupole moment and therefore affect the gravitational coupling in the binary system \citep{Applegate92}. For NN Ser, this mechanism has been excluded on energetic grounds, due to the low mass of the secondary star and the limited energy production \citep{Brinkworth06}. Indeed, one may expect that similar arguments can be made for many systems with low-mass companions \citep{Zorotovic13}. The possibility of apsidal precession has also been excluded based on recent data \citep{Parsons14}, leaving only the planetary hypothesis as a possible explanation.

In this article, we will first review the main arguments against a first-generation origin of the planets given by \citet{Volschow14} in section 2, and subsequently put forward a potential formation scenario for second-generation planets in section 3 \citep[see][]{Schleicher14}. Our main results and conclusions are summarized in section 4.

\section{The case against a first-generation origin}
To assess the feasibility of a first-generation scenario, we need to reconstruct the properties of the system before the common envelope phase. We will here discuss the case of initially spherical orbits, and refer to \citet{Volschow14} for the more general case. Following \citet{Beuermann10}, we expect an initial size of the binary of about $1.44$~AU, corresponding to the typical size of a red giant. The latter allows the system to enter the common envelope phase during the later stages of the evolution. Using the binary star evolution code by \citet{Hurley02}, \citet{Mustill13} have reconstructed the mass of the white dwarf progenitor to be between $1.875$~M$_\odot$ and $2.25$~M$_\odot$, assuming metallicities between $0.01$ and $0.03$. As the current mass of the white dwarf corresponds to $M_{WD}=0.535$~M$_\odot$, we have a mass loss factor $\mu=M_{\rm current}/M_{\rm prev}\sim0.3$. The mass of the secondary corresponds to $M_2=0.111$~M$_\odot$, and plays only a minor role in the mass budget.  In such a system, it is well-known that the planetary orbits are dynamically unstable, unless they exceed at least three Hill radii. As a result, one expects semi-major axis of $r_p\sim2.5$~AU. Assuming Keplerian rotation, we have\begin{equation}
v_p^2 = \frac{GM_{ini}}{r_p},
\end{equation}
where $M_{ini}$ is the initial mass of the system. Due to the mass loss of the system, the gravitational binding energy will decrease, so the orbits of the planets will widen or even become unbound. Considering that the escape velocity is given as\begin{equation}
v_{esc}^2 = \frac{2GM}{r_p},
\end{equation}
with $M=\mu M_{ini}$ the central mass after the ejection, the planets will become unbound for a mass loss factor $\mu<0.5$. As mentioned above, a mass loss of $\mu\sim0.3$ is expected for NN Ser, and it seems likely that the previous planets became gravitationally unbound. Even for more moderate mass losses, one can show that the planetary orbits would evolve as \citep{Volschow14}\begin{equation}
a_f = \frac{r_i}{2-\mu^{-1}},
\end{equation}
with $a_f$ the semi-major axis after the mass loss and $r_i$ the initial orbital radius. Assuming a typical expansion factor of $3$, we therefore expect planetary orbits on scales of $\sim10$~AU, considerably larger than the observed ones.

In these calculations, we have implicitly assumed that the mass loss occurs almost instantaneously. This is plausible, as most of the energy is released in the late stages of the common envelope phase, where the secondary has spiraled down to scales close to the central core, and the energy release occurs on very short timescales \citep[see e.g.][]{Kashi11}. However, the models for the common envelope phase are still highly uncertain, and it is also conceivable that, at least for part of the evolution, the mass loss occurs more gradually. Such a scenario has been explored by \citet{Zwart13} for the planets in HU~Aqua. In this case, the orbits of the planets will gradually adjust according to the decreasing mass of the system. 

From the conservation of angular momentum, one can then show that $r_f = r_i \mu^{-1}$, with $r_f$ and $r_i$ the final and initial radius of the spherical orbit. With the expected mass loss factor $\mu\sim0.3$, we therefore find  an orbital increase by a factor of $\sim3$, suggesting that the observed planetary orbits should be $\sim10$~AU, while in fact the planets have been detected with semi-major axes of $3.3$~AU and $5.4$~AU.

Additional arguments can be made based on the eccentricities of the planets. In the orbital solution of \citet{Beuermann13}, these are given as $\epsilon_1=0.144$ and $\epsilon_2=0.222$ for the outer and inner planet. In the case of an instantaneous mass loss, one would expect a significant increase of the eccentricity as \citep{Volschow14}\begin{equation}
\epsilon_f=\mu^{-1}-1.
\end{equation}
For mass loss factors of $\mu\sim0.5$, we therefore  expect eccentricities of order $1$. Similar results have  been reported by \citet{Mustill13} based on a dynamical analysis.

To avoid these constraints, gas drag forces are necessary to decrease the planetary velocities by friction effects to keep them gravitationally bound and at low eccentricities. For this purpose, \citet{Volschow14} considered both a spherically symmetric outflow assuming an exponential density profile, as well as a disk-shaped ejection mechanism in the orbital plane of the binary system. In the spherically symmetric case, their calculation shows that the drag forces is clearly subdominant, with a typical ratio of $10^{-10}$ compared to the gravitational force. For a disk-shaped outflow, there can be some initial drag inwards, which is  compensated during the subsequent evolution, where the drag forces are reduced during the expansion of the disk, and even an outward acceleration may occur when the disk material moves beyond the planetary orbit. In fact, in most of the simulations the planets are  ejected, and it appears unlikely that the observed low-eccentricity orbits could be produced. A potential way to avoid the problem could be an angular momentum exchange with an inhomogeneous gas distribution, which could however easily produce an infall of the planets onto the star. Alternatively, one could increase the effect of the gas drag and the resulting friction by maintaining high gas densities near the planets for several orbital periods. The latter can be the case if some material is ejected but remains gravitationally bound, leading to the formation of a fall-back disk. In such a case, the gas will  be accreted onto the planets and affect their dynamical evolution. In the next section, we will consider the formation of such a fall-back disk and the formation of new planets via gravitational instabilities. These planets may either form from seeds created by the gravitational instability, or potentially also from the previous planets, which may become the cores of the new ones. In the first case, we expect a truely second generation, while the second case corresponds to a hybrid scenario between first- and second-generation planet formation.

\section{Planet formation from the ejecta of common envelopes}

In this section, we describe the formation of a fall-back disk after the common-envelope phase, as well as the formation of giant planets via gravitational instabilities as outlined by \citet{Schleicher14}. 
\subsection{Ejection and the formation of a fall-back disk}
\begin{figure}[htbp]
\begin{center}
\includegraphics[scale=0.47]{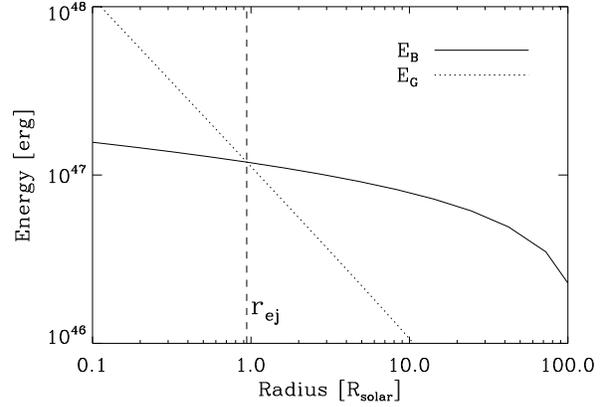}
\caption{Comparison of the gravitational binding energy of the envelope with the gravitational energy released via the inspiral in NN Ser. Both energies become equal for $E_B=E_G\sim1.2\times10^{47}$~erg at a radius of $\sim0.94$~$R_\odot$, leading to the ejection of the envelope \citep{Schleicher14}.}
\label{fig:energy}
\end{center}
\end{figure}

For the ejection event, we adopt the model of \citet{Kashi11} as employed by \citet{Schleicher14}. We assume the AGB star to consist of a core with $M_{\rm core}=0.535$~M$_\odot$ and an envelope with $M_{\rm env}=1.465$~M$_\odot$, consistent with the parameters inferred by \citet{Mustill13}. The mass enclosed within radius $r$ is then given as\begin{equation}
M(r)=M_{\rm core}+\int_{R_{\rm core}}^r 4\pi r^2\rho(r)dr,
\end{equation}
with $R_{\rm core}\sim 0.01$~R$_\odot$. As in the models of \citet{Nordhaus06}, \citet{Tauris01} and \citet{Soker92}, we assume a power-law density profile in the envelope of the AGB star where\begin{equation}
\rho(r)=Ar^{-\omega},
\end{equation}
with $\omega=2$. From the mass of the envelope, the normalization follows as\begin{equation}
A=\frac{M_{\rm env}}{4\pi R_*}
\end{equation}
with $R_*$ the radius of the AGB star, for which we adopt a typical value $R_*\sim185$~R$_\odot$. We consider now the inspiral of the secondary star with mass $M_2=0.111$~M$_\odot$ and assume that the ejection occurs once that the released gravitational energy exceeds the binding energy of the envelope mass outside radius $r$. The latter is given as \begin{equation}
E_B=\int_r^{R_*} \frac{G(M(r)+M_2)}{r}4\pi r^2\rho(r)dr,
\end{equation}
while the former is calculated via\begin{equation}
E_G=\frac{GM(r)M_2}{2r}-\frac{G(M_{\rm core}+M_{\rm env})M_2}{2R_*}.
\end{equation}
In this equation, the factor $1/2$ reflects that half of the gravitational potential is balanced by the kinetic energy. We compare both expressions in Fig.~\ref{fig:energy} and show that they are equal at a radius of $0.9$~R$_\odot$. In general, one may therefore expect the mass in the envelope to be ejected, but one needs to determine which fraction of the mass becomes truely unbound. The latter requires to also determine the velocities of the ejected material. Clearly, the latter can be pursued only in an approximate fashion within an analytical framework \citep[see e.g.][]{Kashi11}, while numerical simulations may not be able to address the later stages of the evolution \citep[cf.][]{Passy12}.

In the following, we will assume that the energy injection occurs almost instantaneously, which may be justified due to the short timescales once that the secondary has reached small scales close to the central core. This is also the stage where most of the gravitational energy will be released. We can therefore employ the self-similar Sedov solution for a power law density profile \citep{Sedov59}. As the injected energy $E_0$, we consider the sum of the binding energy of the envelope $E_B$ and the available thermal energy in the envelope, which we calculate from the virial theorem as $0.5E_B$. The position of the shock front is then expected to evolve as \begin{equation}
R_S(t)=\left( \frac{E_0 t^2}{\alpha A} \right)^{1/(5-\omega)},
\end{equation}
where the constant $\alpha$ can be calculated from energy conservation. It is then possible to obtain the post-shock profiles of the radial velocity, density and pressure as \citep{Kashi11, Schleicher14}\begin{eqnarray}
v&=&\frac{2}{3\gamma-1}\frac{r}{t}=\frac{1}{2}\frac{r}{t},\\
\rho &=&\frac{A(\gamma+1)}{r^\omega(\gamma-1)}\lambda^{8/(\gamma+1)}=4Ar^{-2}\lambda^3,\\
p&=&\frac{A}{r^{\omega-2}t^2}\frac{2(\gamma+1)}{(3\gamma-1)^2}\lambda^{8/(\gamma+1)}=\frac{1}{3}At^{-2}\lambda^3,
\end{eqnarray}
where we assumed $\gamma=5/2$ after the second equality and introduced the self-similar variable \begin{equation}
\lambda=\left( \frac{A\alpha}{E_0} \right)^{1/(5-\omega)}rt^{-2/(5-\omega)}=r/R_S.
\end{equation}
From the consideration that $E_{\rm therm}+E_{\rm kin}=E_0$, one can show that $\alpha=\pi$. We evaluate the velocity profile at the time where $R_S=R_*$, and compare with the escape velocity of the system,\begin{equation}
v_{\rm esc}(r)=\left( \frac{2G(M_{\rm core}+M_2)}{r} \right)^{1/2}.
\end{equation}
As a result, we find that the mass on scales above $106$~R$_\odot$ becomes unbound. The ejected mass which remains bound to the system corresponds to $M_{\rm bound}\sim0.133$~M$_\odot$ or $140$ Jupiter masses. In reality, the bound fraction may even be higher if the ejection process is highly inhomogeneous.

A relevant question  concerns the angular momentum of the ejected material. The total angular momentum that is deposited in the envelope by the secondary can be estimated as \begin{equation}
L_{\rm dep}=M_2 R_*\sqrt{\frac{GM_1}{R_*}},
\end{equation}
yielding about $1.2\times10^{52}$~erg~cm$^2$~s$^{-1}$. It is however likely that the envelope was previously rotating, as angular momentum can be exchanged with the secondary via tidal torques before the onset of the common envelope phase \citep{Hut81, Soker95, Hurley02, Zahn08}. As shown by \citet{Bear10}, the rotational velocity can reach $f_{\rm rot}=45\%$ of the breakup velocity. The angular momentum due to rotation of the envelope is thus given as\begin{equation}
L_{\rm rot}=\int_0^{R_*}f_{\rm rot}\cdot4\pi r^2 dr\rho(r)r\left( \frac{GM(r)}{r}\right)^{1/2},
\end{equation}
yielding a contribution of $L_{\rm rot}\sim2.6\times10^{52}$~erg~cm$^2$~s$^{-1}$. To determine the spatial extent of the disk, the main quantity of interest is however the specific angular momentum, which we parametrize as\begin{equation}
\frac{L_{\rm disk}}{M_{\rm disk}}=\alpha_L\frac{L_{dep}}{M_{ej}}.
\end{equation}
Both due to the inhomogeneous injection of the angular momentum of the secondary, as well as due to the initial rotation of the envelope, one may expect that a value of $\alpha_L\sim10$ can be achieved \citep[see discussion in][]{Schleicher14}. Similar enhancements of the specific angular momentum have been observed for instance in the gaseous disks in the Red Rectangle \citep{Bujarrabal03, Bujarrabal05}. We also note that an independent analysis of the angular momentum constraints has been pursued by \citet{Bear14}, finding no relevant constraints for  planet formation in NN Ser.

\subsection{Planet formation via gravitational instabilities}

\begin{figure}[htbp]
\begin{center}
\includegraphics[scale=0.47]{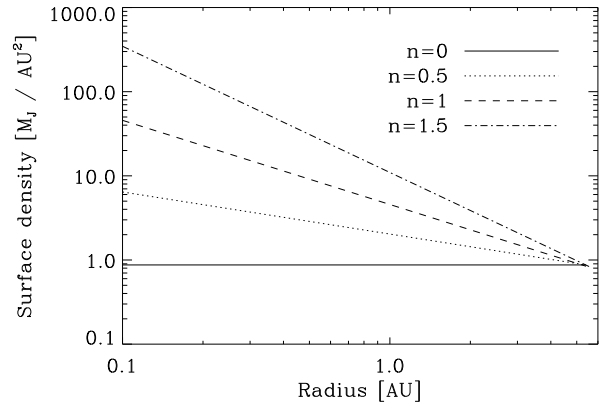}
\caption{{The gas surface density as a function of radius for the models outlined by \citet{Schleicher14}, with power-law indices from $n=0$ to $n=1.5$.The models are normalized to reproduce the observed planetary mass at $5.4$~AU \citep{Schleicher14}.}}
\label{fig:sigma}
\end{center}
\end{figure}

In the following, we assume that the gas of the gravitationally bound material settles into a disk described by a power-law profile of the gas surface density,\begin{equation}
\Sigma(r)=\Sigma_0\left( \frac{r_{out}}{r} \right)^n,
\end{equation}
where $r_{out}$ denotes the outer radius of the disk, $\Sigma_0$ the disk surface density at the outer radius, and $n$ the power-law index. We will in particular consider the case of a Mestel disk with $n=1$ \citep{Mestel63}. We note here that $\Sigma_0$ and $r_{out}$ can be calculated from the disk mass and angular momentum as outlined in \citet{Schleicher14}, and the resulting surface density profiles are given in Fig.~\ref{fig:sigma}.

We expect that the disk efficiently cools through the emission of the dust grains, and the Toomre Q parameter reaches as marginally stable state with\begin{equation}
Q=\frac{c_s\Omega}{\pi G \Sigma}\sim1.
\end{equation}
We therefore have\begin{equation}
c_s=\frac{\pi G\Sigma}{\Omega}.
\end{equation}
We note that $\Omega$ is calculated assuming Kepler rotation with a central mass corresponding to the mass of the present system. The disk height  follows as $h(r)=c_s(r)/\Omega(r)$ \citep{Levin07, Lodato07}.

\begin{figure}[htbp]
\begin{center}
\includegraphics[scale=0.47]{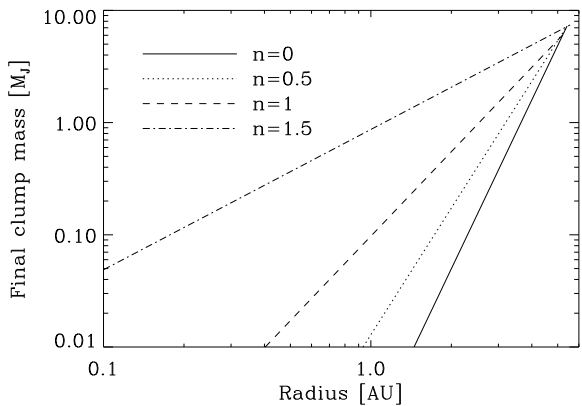}
\caption{{The final clump masses as a function of radius  for the models by \citet{Schleicher14} in the absence of radiative feedback, with power-law slopes from $n=0$ to $n=1.5$ \citep{Schleicher14}. }}
\label{fig:clumpf}
\end{center}
\end{figure}

The mass of the initial clumps forming by the gravitational instability is then given as \citep{Boley10, Meru10, Meru11, Rogers12} \begin{equation}
M_{cl}=\Sigma(r)h^2(r).
\end{equation}
These clumps may grow on the orbital timescale until they reach the gap opening mass where the whole disk is evaporated near their orbit \citep{Lin86}:\begin{equation}
M_{f}=M_{cl}\left[ 12\pi \left( \frac{\alpha_{\rm crit}}{0.3}\right)\right]^{1/2}\left( \frac{r}{h} \right)^{1/2}.
\end{equation}
Here $\alpha_{\rm crit}$ denotes the $\alpha$-parameter for viscous dissipation for which fragmentation occurs. We adopt here a generic value of $\alpha_{\rm crit}\sim0.3$ \citep{Gammie01}. The resulting clump masses are given in Fig.~\ref{fig:clumpf}. We note in particular that it is straightforward to reproduce the characteristic mass scale of the planets, while the expected position has a stronger dependence on the specific angular momentum. In addition, the potential effect of migration has to be considered \citep[e.g.][]{Baruteau11}. As a result of such migration, the observed resonances of the planets \citep{Beuermann13} may form on timescales of a few orbital periods.

\subsection{Radiative feedback}
Radiation from the central star can substantially heat up the disk and stabilize it against gravitational instabilities in the interior. We adopt here the approach of \citet{Chiang97} to estimate the temperature in the midplane of the disk as\begin{equation}
T=\left( \frac{\theta}{4} \right)^{1/4}\left( \frac{r_*}{r} \right)^{1/2}T_*,\label{temp}
\end{equation}
where $T_*$ is the effective temperature of the central star and $r_*$ its radius. We consider here the stage very early after the ejection, and therefore estimate the radius of the star to be comparable to the remaining radius of the core \citep[see e.g.][]{Schleicher14}. The grazing angle $\theta$ describes the angle at which light from the star strikes the disk, and is estimated as $\theta\sim0.4 r_*/r$. For the temperature, we then adopt the minimum of Eq.~\ref{temp} and the temperature expected in a marginally stable state ($Q\sim1$). 

\begin{figure}[htbp]
\begin{center}
\includegraphics[scale=0.47]{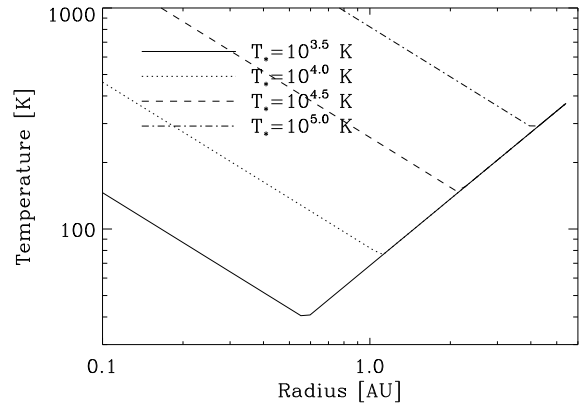}
\caption{{The gas temperature in the presence of photoheating from the star in the case of a Mestel disk, corresponding to model C described by \citet{Schleicher14}. We explore the effect of different effective temperature of the central star $T_*$. The minimum temperature in the model is based on the assumption that the disk should be marginally stable \citep{Schleicher14}. }}
\label{fig:temp_rad}
\end{center}
\end{figure}

The resulting thermal evolution is given in Fig.~\ref{fig:temp_rad}, showing that radiative feedback can stabilize the disk on scales around a few AU. The latter may naturally explain the stability of the disk in the interior, and could in fact provide a natural point to stop the migration, as it becomes easier to create a gap in the disk once it is gravitationally stable \citep{Baruteau11}.

\subsection{Comparison with the observed population}
While NN Ser is currently the system with the most detailed data and the highest-quality fits to the planetary orbits, \citet{Zorotovic13} have suggested $12$ PCEB-systems which may harbor such massive planets, some of which even indicate the presence of two planets as in NN Ser. These systems include five detached systems with a hot subdwarf B (sdB), four detached systems with a white dwarf (WD) and three cataclysmic variables (CVs). For the system NSVS14256825, the two-planet solution was shown to be dynamically unstable \citep{Wittenmyer13}, while the one-planet solution is stable \citep{Beuermann12b}. In this case, we therefore use the data of \citet{Beuermann12b}, while we adopt the parameters inferred by \citet{Zorotovic13} in the other cases. 

\begin{figure}[htbp]
\begin{center}
\includegraphics[scale=0.47]{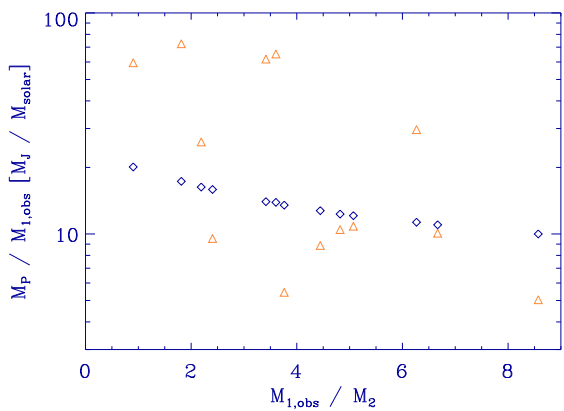}
\caption{Comparison of model predictons (diamonds) with planetary masses inferred from observations by \citet{Zorotovic13} (triangles). The data indicate the potential presence of two populations, one comparable to our model predictions, and one with even higher masses \citep{Schleicher14}.}
\label{fig:stat}
\end{center}
\end{figure}

The comparison of the theoretical predictions with the observationally inferred masses is given in Fig.~\ref{fig:stat}. We find that there is a population showing good agreement with the theoretical predictions, as well as an additional population with even higher masses. The latter may hint either at a different origin of these planets, or the potential presence of additional effects related to magnetic activity, which cannot be ruled out in some cases.

\section{Discussion and conclusions}
In this article, we have summarized the main arguments against a first-generation scenario for the planets in NN Ser, and presented a theoretical model explaining the formation of planets from the ejecta during the common envelope phase. The model naturally explains the observed planetary masses in NN Ser, and is in good agreement with a number of additional systems. In addition, there seems to be a population where the observationally inferred masses significantly exceed the theoretical predictions. It is interesting to note that in exactly these cases, it is difficult to justify a second-generation scenario due to angular momentum constraints \citep{Bear14}. The latter can potentially hint at a different planetary origin. We note that beyond a purely first- or second-generation origin, also hybrid scenarios are conceivable where the existing planets accrete additional mass from the bound material remaining after the ejection. 

The scenario proposed here can be validated through the further observation of the eclipsing binary systems, as a strict periodicity is expected in the case of planetary orbits. In addition, complementary approaches like the detection of the planets via their thermal emission could provide an important pathway to confirm their existence. Telescopes like Gemini and ALMA may in addition look for remants of the gravitationally bound gas in NN Ser to identify a potential debris disk or the gas ejected during the common envelope event. The latter may provide further information to significantly refine such a model.

\acknowledgements
We thank  Klaus Beuermann, Christiane Diehl, Tim Lichtenberg and Sonja Schuh for stimulating discussions on the topic. DRGS thanks for funding from the Volkswagen foundation via the project ''Planets beyond the main sequence: theory and observations''. SD thanks the {\em Deutsche Forschungsgemeinschaft} (DFG) for funding via the SFB~963/1 ''Astrophysical flow instabilities and turbulence'' (project A5).



\end{document}